\documentclass{icrc2009}

\usepackage{graphicx}   
\usepackage[caption=false]{caption}    
\usepackage[font=footnotesize]{subfig} 
\usepackage{fixltx2e}
\usepackage{url}

\newcommand{\shorttitle}[1]%
{\markboth{Proceedings of the 31\MakeLowercase{$^{st}$} ICRC, {\L}\'{o}d\'{z} 2009}{#1} }
\newcommand{\etal}{\MakeLowercase{\textit{et al. }}} 

\begin{document}
\title{Constraining cosmological parameters from the EBL attenuation of VHE spectra of extragalactic sources}

\author{\IEEEauthorblockN{Bagmeet Behera\IEEEauthorrefmark{1}\IEEEauthorrefmark{2} and
			  Stefan J. Wagner\IEEEauthorrefmark{1}}
                            \\
\IEEEauthorblockA{\IEEEauthorrefmark{1}Landessternwarte, Zentrum f\"ur Astronomie der Universit\"at Heidelberg, D-69117, Germany}
\IEEEauthorblockA{\IEEEauthorrefmark{2}Fellow of the International Max Planck Research School for Astronomy and Cosmic Physics\\
at the University of Heidelberg}
}

\shorttitle{Behera \etal Cosmological parameters from EBL}
\maketitle

\begin{abstract}
The Extragalactic Background Light (EBL) causes attenuation of the VHE spectra of extragalactic sources, via pair-production. The amount of attenuation depends on the distance of the source, i.e. on the integration over the line of sight path taken by the VHE photons. The line of sight path calculation, depends on the values of the cosmological parameters and the evolution of these parameters. With the recent indications of a lower EBL level from observations and the greater sensitivity of the proposed future Cherenkov experiments it will be possible to detect higher redshift sources. We study whether the EBL attenuation of extragalactic sources is sensitive enough to put constrains on the cosmological parameters.\\
\end{abstract}

\begin{IEEEkeywords}
VHE, EBL, Cosmology
\end{IEEEkeywords}
 
\section{Introduction}
Measurement of cosmological parameters traditionally involves two common strategies, viz. using standard candles as cosmological probes, and via precision measurements of the cosmic microwave background (CMB). Recent results using a combination of such strategies has succeeded in measuring the values of cosmological parameters with a high degree of precision. The attenuation of the VHE spectra of extragalactic sources can be an independent method to constrain some cosmological parameters such as the Hubble constant (H$_o$), and the cosmological density parameters for matter ($\Omega_m$) and dark energy ($\Omega_\Lambda$).\\

The EBL in the UV to mid-IR regime is responsible for causing attenuation of the VHE gamma-ray flux from extragalactic sources.  The optical depths to $\gamma$-rays is dependent on the energy of the $\gamma$-ray, the  details of the EBL photon density and the distance of the source. For sources at non-negligible redshifts (z $\geq$ 0.1), the changes in the comoving photon density which directly influences the attenuation-level, depends on cosmology. This is due to evolution of the Hubble constant with z, which determines the comoving volume of the universe. as well as the evolution of the EBL density itself. The later can be indirectly measured by making some reasonable assumptions on the intrinsic spectra of sources, and the evolution of the sources contributing to the EBL photon-density, namely the star formation history. In \cite{BlanchMartinez2005} the possibility of using the gamma-ray horizon as a measurement tool for probing cosmology was explored. Since the gamma-ray horizon varies with redshift, its measurement will be plagued due to the varying range of uncertainties in the systematics of the EBL density and the difficulty to measure the gamma-ray-horizon accurately at low redshifts. We present a different parameter to circumvent these issues.\\

Blazars and $\gamma$-ray bursts would be two such classes of extragalactic objects that can be used as probes for EBL absorption. In this work, the attenuation (for a fixed $\gamma$-ray energy, E$_o$) as a function of z, $\eta(E_o, z)$\,$\equiv$\,exp$[-\tau(E_o, z)]$ (where $\tau$ is the optical depth to a $\gamma$-ray of observed energy $E_o$, for a source redshift of $z$), is chosen as the relevant parameter to measure the effect of cosmological parameters. An EBL model based on upper limits derived from VHE observations is adopted to calculate the attenuation. The range of variation in $\eta(E_o, z)$ within the uncertainties of the cosmological parameters, and its sensitivity to measure these variations is estimated.\\

\section{Attenuation of VHE blazar spectra}
  \begin{figure*}[!t]
   \centerline{\subfloat[Attenuation for various z]{\includegraphics[width=2.6in]{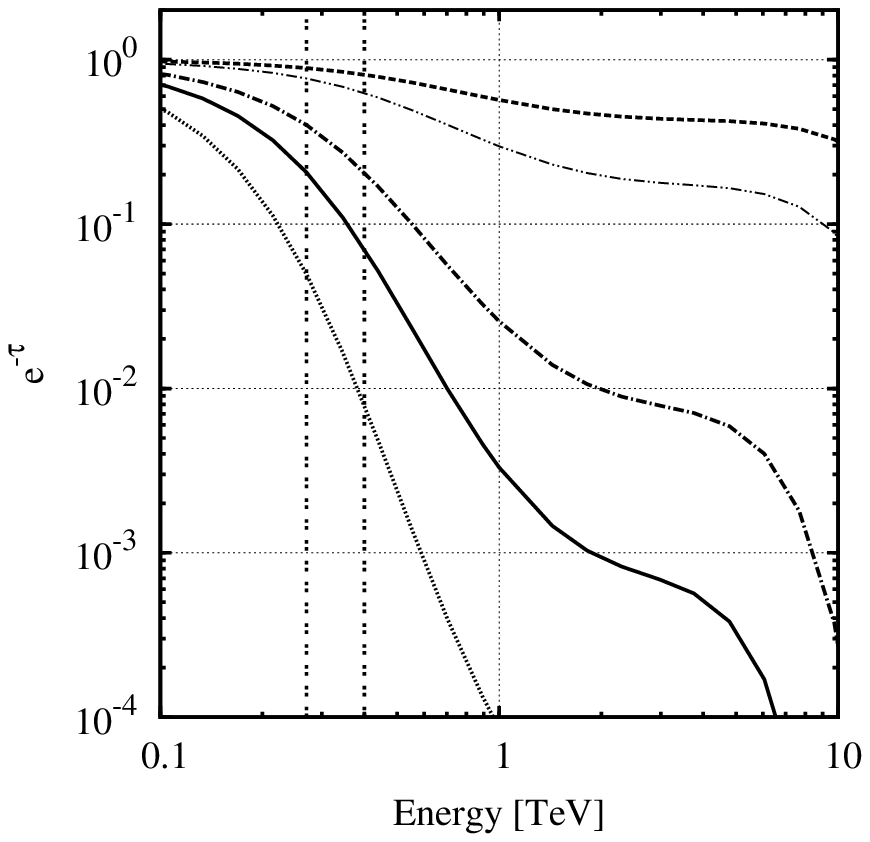}
	  \label{AttenuationVariousEn}}
              \hfil
              \subfloat[Different EBL models]{\includegraphics[width=2.9in]{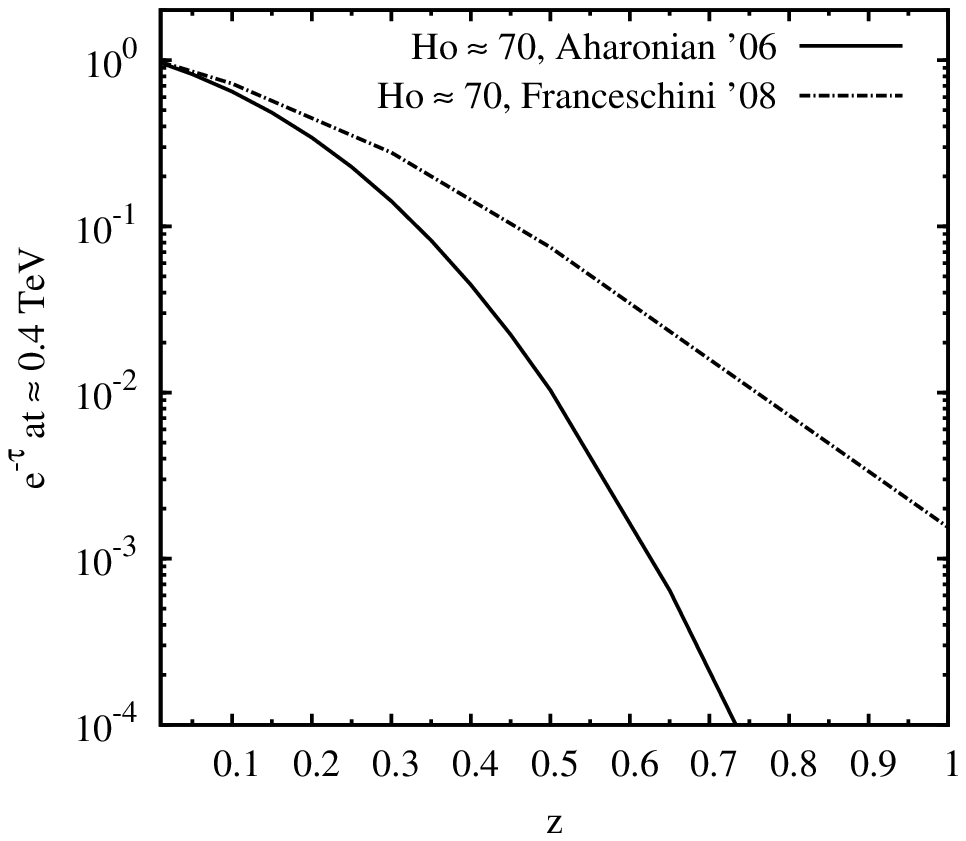}
	  \label{AttenuationVariousEBLmodels}}
             }
   \caption{\textbf{\emph{left:}} The attenuation as a function of the observed $\gamma$-ray energy for a number of redshifts. Top-most dashed curve is for z = 0.05, below which are curves for z = 0.1, 0.25, 0.35 and 0.5. \textbf{\emph{right:}} Comparison of two different EBL models. The solid line is for the EBL upper limits derived from observations of VHE-blazars (used in the work presented here), and the dashed-dotted line is for an EBL model derived from theoretical models combined with direct source (contributing to the EBL) observations.}
   \label{Attenuation}
\end{figure*}
VHE $\gamma$-rays are absorbed in the inter-galactic medium via pair-production mechanism, from photon-photon scattering on the extragalactic photon field. The optical depth due to pair production from the $\gamma_{\scriptstyle CIB}+\gamma_{\scriptstyle VHE }\rightarrow e^-+e^-$ interaction is given by:
\[\tau(E_o,z_s) = \sigma_T\int_{0}^{z_s} dz \frac{dl}{dz} \int_{\epsilon_{min}}^{\epsilon_{max}} d\epsilon \{(1+z)^3\times\]
\begin{equation}
\label{eq:Tau_eqn}
\quad\quad\quad\quad n_{\epsilon}[\frac{\epsilon}{1+z}]\sigma_{\gamma\gamma}[s=2E_o (1+z) \epsilon]\}
\end{equation} where $E_o$ is the observed $\gamma$-ray photon energy from blazar at $z=z_s$, $n_\epsilon(z)$ is the comoving EBL density, $\sigma_{\gamma\gamma}[s]$ is the $\gamma-\gamma$ interaction cross-section (integrated over all angles) as a function of the center of mass energy, and $\sigma_T$ is the Thomson cross-section. The cosmological parameters go into the expression $dl/dz$ as:
\begin{equation}
\label{eq:dlOdz}
 \frac{dl}{dz} = \frac{c}{H_o (1+z) \sqrt{\Omega_m (1+z)^3 + \Omega_\lambda}}
\end{equation} where $H_o$ is the Hubble constant at $z$\,$=$\,$0$ and $\Omega_m$, $\Omega_\Lambda$ are the matter and dark-energy density parameters respectively. This expression is assuming $\Lambda$CDM cosmology, and neglecting the radiation density $\Omega_r$.\\

To derive constrains on the parameters $H_o$, $\Omega_m$ and $\Omega_\Lambda$, it is necessary to have a estimate on the uncertainty in the EBL density and the intrinsic blazar spectra. EBL models are estimated either by theoretical methods, such as \cite{SalamonStecker1998, Primack2005, Kneiske2004, Franceschini2008}(the last one is based on measurements and theoretical assumptions) or using TeV observations such as \cite{Aharonian2006a}. A set of attenuation curves for sources at various redshifts is shown in figure \ref{AttenuationVariousEn} for the EBL model taken from \cite{Aharonian2006a}. This EBL model is based on reasonable assumptions on the intrinsic spectrum of blazars and are realistic upper limits on the EBL photon density. The EBL model in \cite{Franceschini2008} on the other hand can be considered close to a lower limit. A comparison of the attenuation for a $0.4$\,TeV $\gamma$-ray photon from various source redshifts calculated using these two EBL models is shown in figure \ref{AttenuationVariousEBLmodels}. This can be considered as the uncertainty in the parameterization of the EBL density. To derive meaningful cosmological constrains this uncertainty in the EBL has to be resolved eventually. Apart from direct measurements that are problematic due to strong foreground contamination (see \cite{Hauser2001} for review) the only viable method is through VHE observation of more blazars at higher redshifts. This is a definite possibility in the near future since new IACT instruments are currently under construction (H.E.S.S. II, and MAGIC II) that would increase the sensitivity in the VHE regime and enable us to probe deeper in z. Future experiments like CTA and AGIS will further increase this capability. This combined with Fermi-GST will help constrain the intrinsic spectrum of blazars. Under this assumption we can consider any one EBL model as a prototype to study possible constrains on the cosmological parameters. Since we expect the EBL density to be resolved with VHE observations we take the current upper limits derived in \cite{Aharonian2006a} as the prototype for this study.\\
\begin{figure*}[!t]
   \centerline{\subfloat[H$_o$]{\includegraphics[width=2.75in]{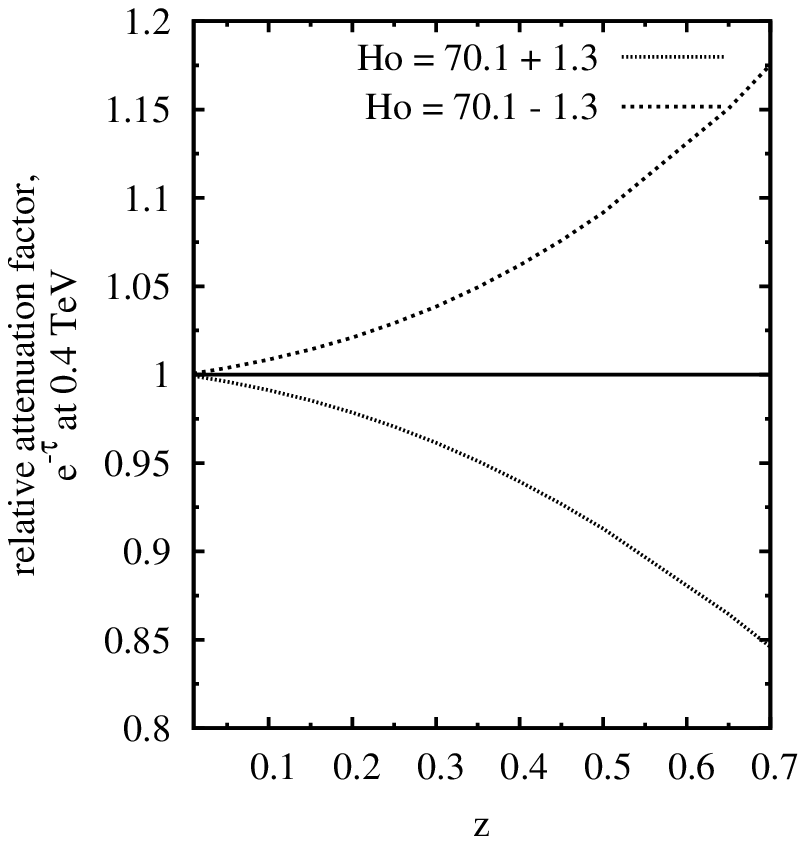}
	  \label{Ho}}
              \hfil
              \subfloat[$\Omega_{m}$]{\includegraphics[width=2.75in]{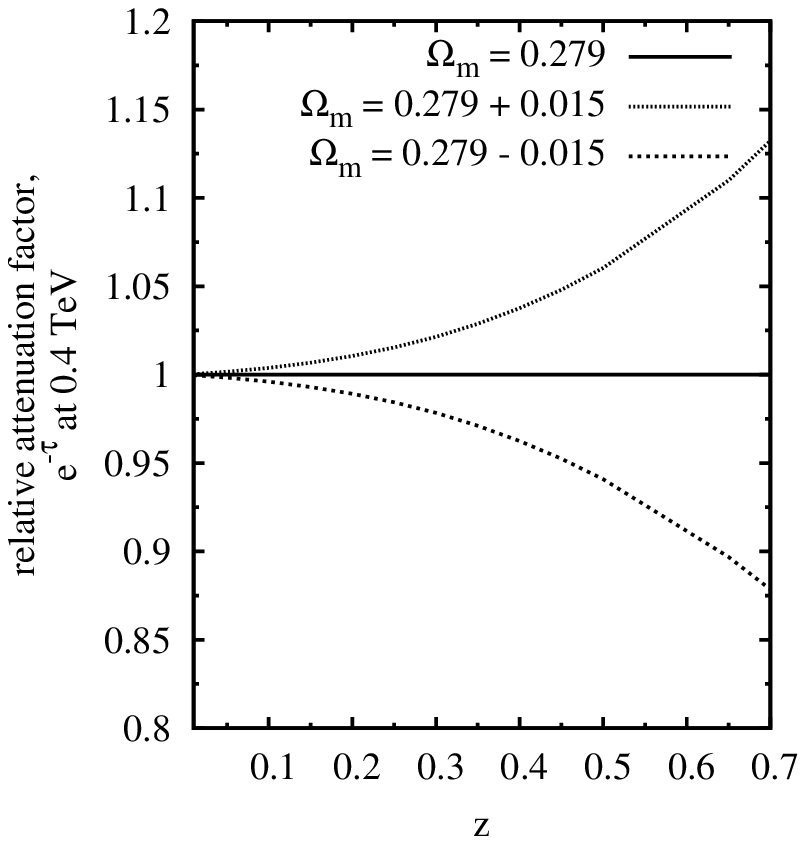}
	  \label{Omegam}}
             }
    \centerline{\subfloat[$\Omega_{\Lambda}$]{\includegraphics[width=2.75in]{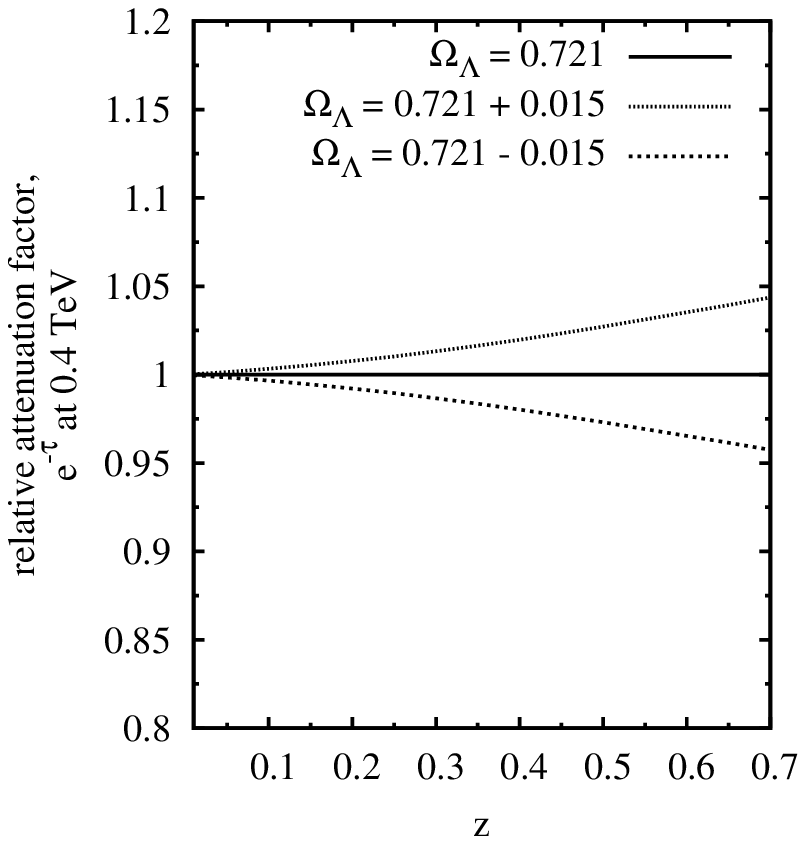}
	  \label{Omegal}}
              \hfil
              \subfloat[Sensitivity]{\includegraphics[width=2.75in]{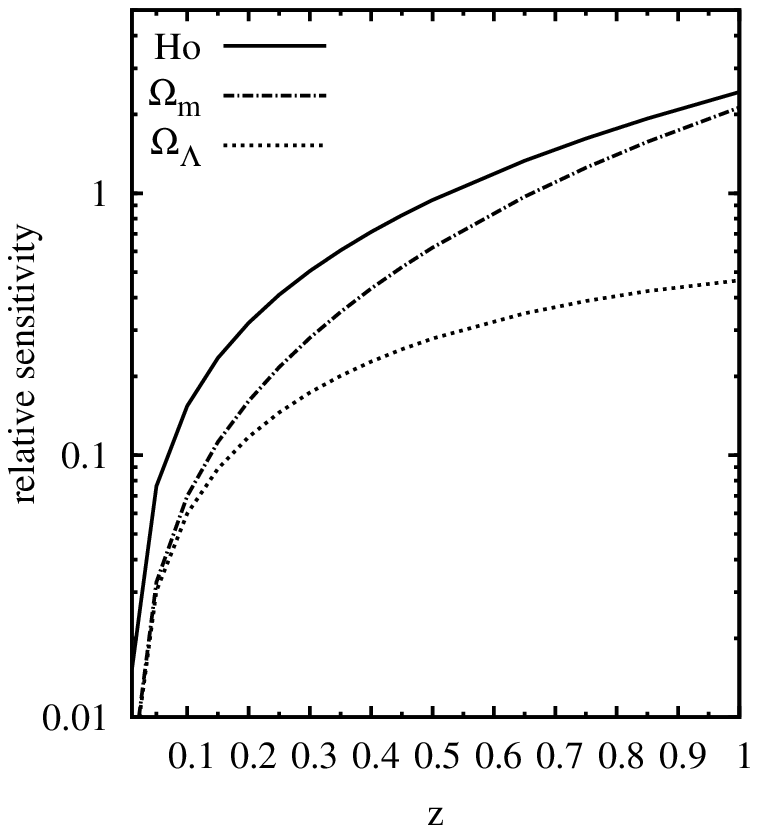}
	  \label{Sensitivity}}
             }
   \caption{The change in attenuation factor with redshift (for a particular cosmology, defined by the three cosmological parameters used here), relative to the cosmology defined by H$_o$ = $70.1$ km/s/Mpc, $\Omega_{m}$ = 0.279, and $\Omega_{\Lambda}$ = 0.721 (horizontal line). The effect of variation due to errors in the known value of:   H$_o$ (\textbf{\emph{top-left}}), $\Omega_{m}$(\textbf{\emph{top-right}}), and $\Omega_{\Lambda}$(\textbf{\emph{bottom-left}}). \textbf{\emph{Bottom-right:}} The relative sensitivity, `s' (see text) of the parameter $\eta(0.4 TeV, z)$, to changes in each cosmological parameter, with redshift is shown.}
   \label{Results}
\end{figure*}

To proceed with our estimate the measurable parameter that should be used as the proxy to probe cosmology needs to be established. This parameter will obviously be a function of the attenuation, which is a directly measurable quantity. The criteria that this proxy-parameter should satisfy are the following: (1) it should have good dynamic range, (2) should also be clearly measurably over a sufficiently large z (z$\sim0.5$), and (3) it should be least affected by systematic uncertainties in the EBL density as well as intrinsic VHE parameters. as per the second criteria - since blazar spectra will get steeper with z (due to EBL absorption), this parameter should be measurable with IACTs at low energies ($<\sim1TeV$) where the instruments are most sensitive. Furthermore, this is also the regime which will put the strongest constrain on the EBL density for the same reason. To make sure that we get a good dynamic range, energies lower than $\approx 0.25 TeV$ should be avoided (see figure \ref{AttenuationVariousEBLmodels}). This rules out using the gamma-ray horizon since it easily exceeds this range for both low and high redshifts. The attenuation measured at $0.4$\,TeV, $\eta(0.4 TeV, z)$ was thus chosen as the appropriate proxy since it satisfies all our criteria. Furthermore, the attenuation between $0.27$\,TeV and $1$\,TeV is approximately linear in log-log representation, thus a powerlaw intrinsic spectra will produce a powerlaw observed spectra. This will be useful to reduce the systematics on the intrinsic spectral parameters (such as the hardest possible intrinsic-spectra, and turn-offs), by considering a  large VHE-sample in the future.\\

\section{Probing cosmology}
To test the sensitivity of $\eta(0.4 TeV, z)$ within the range of uncertainties in the parameters, $H_o$, $\Omega_m$ and $\Omega_\Lambda$, that we want to constrain - we calculated the dispersion in $\eta(0.4 TeV, z)$ within these uncertainties. We assumed that there is no uncertainty on the EBL density, and the measurement error on $\eta(0.4 TeV, z)$ is very small. To this end, one of the three cosmological parameters was alternately varied by the $3 \sigma$ uncertainty from measurements from WMAP5 (combined with two other measurements, taken from \url{http://lambda.gsfc.nasa.gov/product/map/current/params/lcdm_sz_lens_wmap5_bao_snall.cfm}), while keeping the other two at the measured value. The resulting dispersion in $\eta(0.4 TeV, z)$ scaled to the $\eta(0.4 TeV, z)$ calculated at the measured values of all three parameters ($H_o = 70.1$ km/s/Mpc, $\Omega_m = 0.279$ and $\Omega_\Lambda = 0.721$) is shown in figures \ref{Ho}, \ref{Omegal} and \ref{Omegam}. As seen, $H_o$ and $\Omega_m$ show a bigger effect on $\eta(0.4 TeV, z)$, than $\Omega_\Lambda$, with $\approx$ $10\%$, $5\%$ and $2.5\%$ effect at a redshift of $0.5$ for the three parameters respectively.\\

To account for a measurement uncertainty on $\eta(0.4 TeV, z)$, we defined a relative sensitivity parameter $s(p)$ as a function of the fractional change in $\eta(0.4 TeV, z)$ times the inverse of the error on the measurement of $\eta(0.4 TeV, z)$, i.e.\\
\begin{equation}
\label{eq:s(p)}
s(p) =\frac{\Delta \eta_{\delta p}(0.4 TeV, z)}{\eta_{\delta p}(0.4 TeV, z)} \times \left[\eta_{Err}(0.4 TeV, z)\right]^{-1}
\end{equation} - where $\Delta \eta_{\delta p}(0.4 TeV, z) = \eta_{p + \delta p}(0.4 TeV, z) - \eta_p(0.4 TeV, z)$; $\eta_{Err}(0.4 TeV, z)$ is the measurement error which we varied linearly between the arbitrarily chosen  values of $5\%$ at $z = 0.05$ to $15\%$ at $z=1.0$. This was based on the assumption that higher attenuation at higher z, would lower the observed flux and hence the error on the measured flux (and thus the attenuation factor) will also increase with z.\\

We see that the sensitivity for the various parameters vary similarly with z, though to a different degree in all three. Whether the change in sensitivity of these parameters with z, can be used to constrain each other needs more careful study, left out in this work. It should however, be noted that the sensitivity increases exponentially with redshift, hence has the potential to probe the cosmology to the largest redshift that can be reached with IACT experiments.\\

\section{Conclusion}
A study was made to explore the possibility of constraining cosmological parameters with the attenuation of the VHE spectra of blazars. Under the assumption -\newpage \noindent that in the near future the uncertainties in the EBL densities and the intrinsic blazar spectra can be minimized, results indicate that there is a possibility to constrain the three cosmological parameters viz. $H_o$, $\Omega_m$ and $\Omega_\Lambda$, using an accurate measurement on the spectral-attenuation at a fixed energy. This novel parameter is used for the first time in such a study, and has the advantage that its sensitivity to the cosmological parameters probed, increases exponentially with redshift. The uncertainties involved in these measurements is a subject of further study and will be presented elsewhere. It is stressed that this method would provide constrains on these cosmological parameters independent of the traditional methods.\\

\end{document}